\documentclass[twocolumn]{webofc}

\usepackage[varg]{txfonts}   
\usepackage{hyperref}
\usepackage{url}
\usepackage{float}
\hypersetup{colorlinks=true,citecolor=blue,urlcolor=blue,linkcolor=blue}

\begin{document}

\title{Study of $0^+$ and $8^-$ states in even-even $^{250-260}$No isotopes}

\author{\firstname{M.A.} \lastname{Mardyban}\inst{1,}\inst{2}\fnsep\thanks{\email{mmardyban@theor.jinr.ru}}
\and
\firstname{V.O.} \lastname{Nesterenko}\inst{1,}\inst{2}
}

\institute{Laboratory of Theoretical Physics, Joint Institute for Nuclear Research,
141980 Dubna, Moscow region, Russia
\and
   Federal State University "Dubna", Dubna, Moscow Region, 141980, Russia}

\abstract{
Low-lying $K^{\pi}=0^+$ and isomeric $8^-$ states in even-even  isotopes $^{250-260}$No are explored within
the Quasiparticle Random-Phase Approximation (QRPA) method with Skyrme parametrization SLy4.
The deformations, single-particle (s-p) spectra and pairing in the isotopes are inspected. The
calculations predict a pronounced minimum in the neutron pairing at $A$=252, 254,
which significantly affects the properties of $0^+$ and $8^-$ states  and leads to a correlation of
their spectra. It is shown that $8^-$ isomers are basically  low-energy two-quasiparticle (2qp)
states. The appearance or absence of these isomers in $^{250-260}$No is explained as a combined effect
of the s-p spectra and pairing.
The collective $0^+$ states are predicted in all the isotopes as the lowest multipole non-rotational
excitations. These states are interpreted as a superposition of pairing vibrations and $\beta$-vibrations.
The results are in a reasonable agreement with available experimental data for $^{252,254}$No.}

\maketitle


\section{Introduction}

Even-even nobelium isotopes represent a unique laboratory
to study s-p spectra and pairing in heavy nuclei, which can be useful for
understanding the features of neighboring superheavy nuclei. Among transfermium nuclei, nobelium
isotopes have rather extensive experimental spectroscopic data~\cite{Herz,Hess,Herz2,Hess_EPJA10,nndc}.
Particular interest is focused on isomeric states with K$^\pi$=8$^-$, whose existence
is directly related to the specific arrangement of single-particle orbitals near the Fermi
surface~\cite{Herz,Hess,Herz2}. Quite recently new measurements for $8^-$ and $0^+$ excitations
in $^{254}$No were reported~\cite{Tez22,Forge,Forge2,Wahid}. However,
despite a large experimental  and theoretical effort, the assignment of $8^-$ isomer  and physical
treatment of excited $0^+$  state in $^{254}$No are still disputed, see discussion~\cite{Nest-Mard}.

In this work we present a comparative analysis of the lowest excited $0^+$  and isomeric $8^-$ states
in $^{250-260}$No.
The calculations are performed within fully self-consistent QRPA model~\cite{Rep1,Rep2,Rep3} using Skyrme
force SLy4~\cite{SLy4}. In our previous systematic study~\cite{Nest-Mard}, this force was shown to be relevant
for description of ground-state properties and low-energy spectra in  $^{250-260}$No. The s-p spectra
and pairing characteristics are calculated with the code SKYAX~\cite{Skyax} using a two-dimensional grid
in cylindrical coordinates. Equilibrium deformations are obtained by minimization of the system energy. The
pairing is treated within BCS approach. Details of the calculations can be found elsewhere~\cite{Nest-Mard}.

\section{Results and Discussion}
\subsection{Deformations, pairing, mean field}

In Table~\ref{table:deform}, the calculated axial deformations and pairing gaps are presented.
The isotopes demonstrate a similar quadrupole deformation $\beta_2 \approx 0.30$. There is also
a significant hexadecapole deformation $\beta_4$ which decreases with the mass number $A$.

\begin{table}[h]
\caption{Axial quadrupole  $\beta_2$  and hexadecapole $\beta_4$ deformations as well as proton $\Delta_p$
and neutron $\Delta_n$ pairing gaps in $^{250-260}$No, calculated with the force SLy4.}
\begin{center}
\begin{tabular}{|c|c|c|c|c|c|c|}
\hline
 & $^{250}$No & $^{252}$No & $^{254}$No & $^{256}$No & $^{258}$No & $^{260}$No \\
\hline
$\beta_2$  & 0.304 & 0.303 & 0.304 & 0.303 & 0.298 & 0.299\\
$\beta_4$ & 0.096 & 0.078 & 0.061 & 0.044 & 0.026 &0.014 \\
\hline
$\Delta_p$  & 0.49 & 0.48 & 0.47 & 0.46 & 0.49 & 0.54\\
$\Delta_n$  & 0.51 & 0.32 & 0.28 & 0.45 & 0.44 & 0.46\\
\hline
\end{tabular}
\label{table:deform}
\end{center}
\end{table}

\begin{figure}[h!] 
\centering
\includegraphics[width=8.5cm]{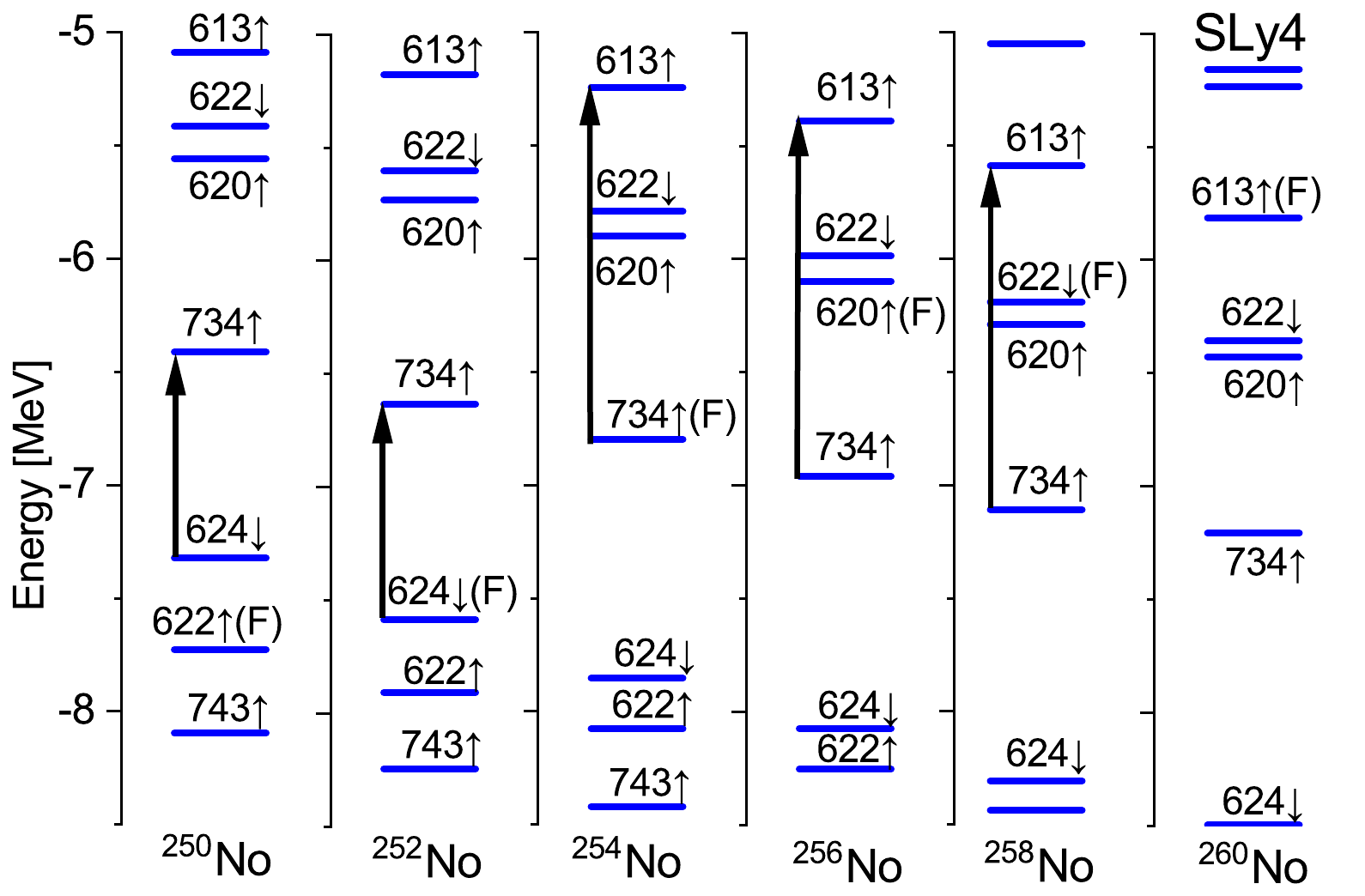}
\caption{Neutron s-p spectra near Fermi levels in $^{250-260}$No. Transitions resulting in  $8^-$ states
(see Table~\ref{tab:8-})  are marked by black arrows. Fermi levels are labeled by (F).}
\label{sps_n}
\end{figure}

As seen from Table~\ref{table:deform}, the proton pairing gaps $\Delta_p$ are rather large and
almost the same for all the isotopes.
Instead, the neutron pairing gaps $\Delta_n$  exhibit a minimum in $^{252,254}$No, caused by the deformation
shell gap in the neutron s-p spectra of these isotopes, see Fig.~\ref{sps_n} and detailed discussion~\cite{Nest-Mard}.
A similar result was previously obtained  with Woods-Saxon potential~\cite{Rob} and various self-consistent
single-particle potentials~\cite{Dobaczewski}. As shown in Fig.~\ref{sps_n},  Fermi levels $624\downarrow$
($^{252}$No) and $734\uparrow$ ($^{254}$No) lie near  or inside the large energy gap $\sim$ 1.7 MeV
in the neutron s-p spectra. Instead, following Fig.~\ref{sps_p}, the proton spectrum near the Fermi level
is rather dense.

\begin{figure}[h!] 
\centering
\includegraphics[width=8.5cm]{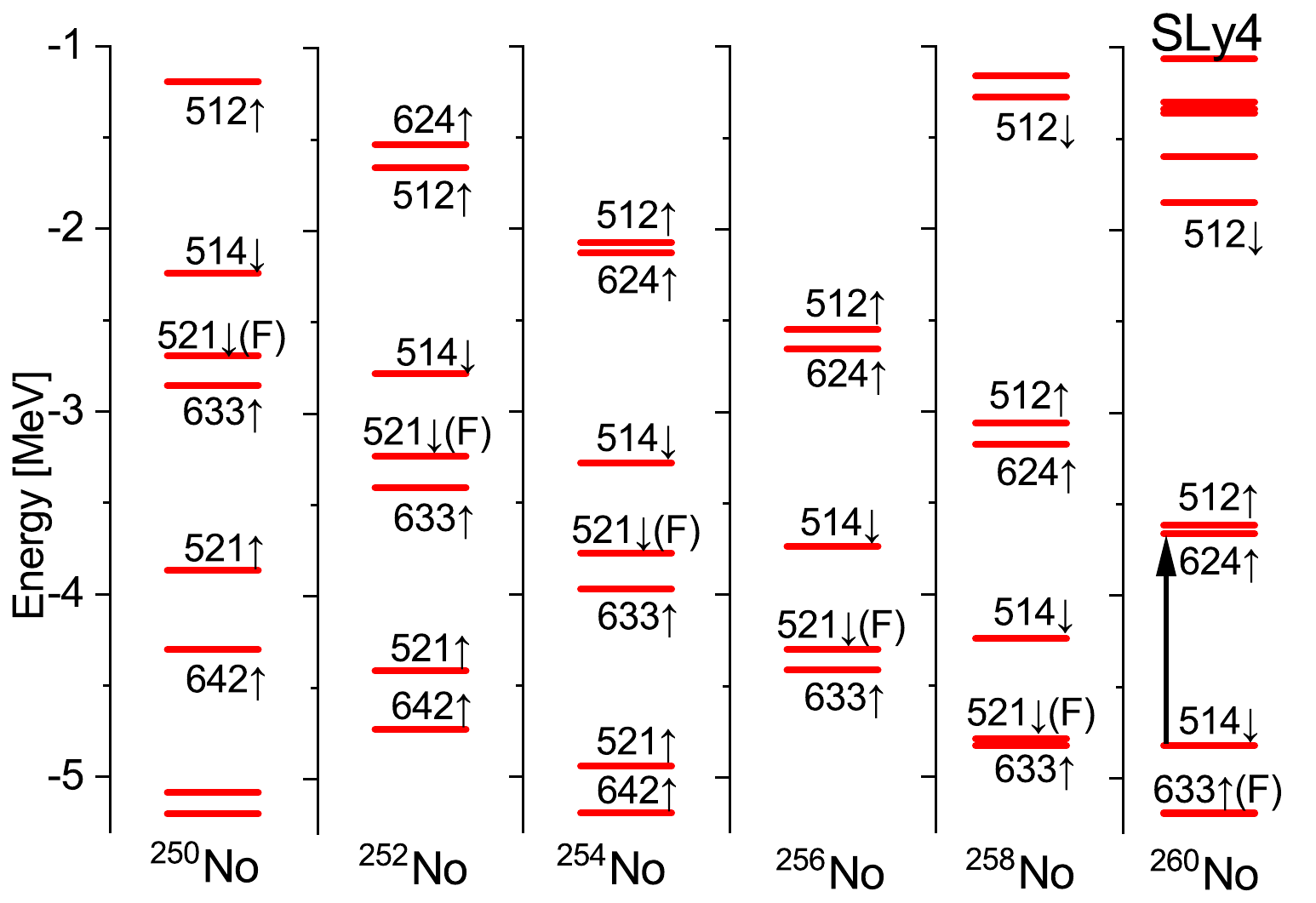}
\caption{The same as in Fig.~\ref{sps_n} but for the proton s-p spectra in  $^{250-260}$No.}
\label{sps_p}
\end{figure}

\subsection{$8^-$ isomers}

The $8^-$ isomers were observed only in two nobelium isotopes: $^{252}$No ($E_x$ =1.254 MeV) and
$^{254}$No  ($E_x$ =1.295 MeV)~\cite{nndc, Herz,Hess,Herz2}. The  $8^-$  isomer in $^{252}$No is
firmly assigned as neutron 2qp excitation $nn[624\downarrow, 734\uparrow]$~\cite{Hess,Nest-Mard}.
The assignment of $8^-$ isomer in $^{254}$No is yet disputed, see the extensive
discussion~\cite{Nest-Mard}. Depending on the model~\cite{Clark,Wahid,Jolos,He,Sol-Sush} and
suggested decay scheme~\cite{Hess_EPJA10,Clark,Forge2,Wahid}, this isomer is treated
as proton pp$[514\downarrow,624\uparrow]$ \cite{Hess_EPJA10,Wahid,Forge2,Jolos,He} or neutron
nn$[734\uparrow,613\uparrow]$~\cite{Clark,Nest-Mard,Sol-Sush}
2qp configuration. Different assignments are caused  by similar excitation energies of these
two configurations, see the estimations~\cite{Nest-Mard,Jolos, Sol-Sush}. The assignment choice
usually depends on a subtle theoretical result -  which configuration,
proton or neutron, is somewhat lower. Even the decay schemes are treated by a contradictive way,
see details in Ref.~\cite{Nest-Mard}.

As shown in Table 2
and Ref.~\cite{Nest-Mard}, our SLy4 calculations support the
neutron assignment $nn[624\downarrow, 734\uparrow]$. This  is consistent with
$\gamma$-decay from a higher-lying $10^+$ isomer to $8^-$ isomer~\cite{Clark}.
To establish the true assignment, the measurement of
gyromagnetic factor $g_K$ (which is essentially different for proton and neutron
states~\cite{BM2,Sol76}) for $8^-$  isomer in $^{254}$No is desirable.

The characteristics of the lowest QRPA $8^-$ states in $^{250-260}$No are shown in Table 2.
All the states are almost pure 2qp configurations. Their contribution to the state norm is
$N_{qq'} \sim$1 and their QRPA energies $E$ are close to 2qp energies $\epsilon_{qq'}$.
A large difference  between 2qp excitation energies $\epsilon_{qq'}$ and energies $e_{qq'}$
of the corresponding transitions between s-p states (exhibited by arrows in Figs.~\ref{sps_n}
and~\ref{sps_p}) denotes a significant pairing impact. In $^{250}$No and $^{260}$No, the relevant
transitions are of particle-particle character and so should be suppressed. In $^{252-258}$No,
$8^-$ isomers are  produced by neutron particle-hole
transitions $624\downarrow \to 734\uparrow$  and
$734\uparrow \to 613\uparrow$. So, following our calculations,
the relevant candidates for low-energy $8^-$ isomers are $^{252-258}$No.

\begin{figure}[h!]
\centering
\includegraphics[width=8cm]{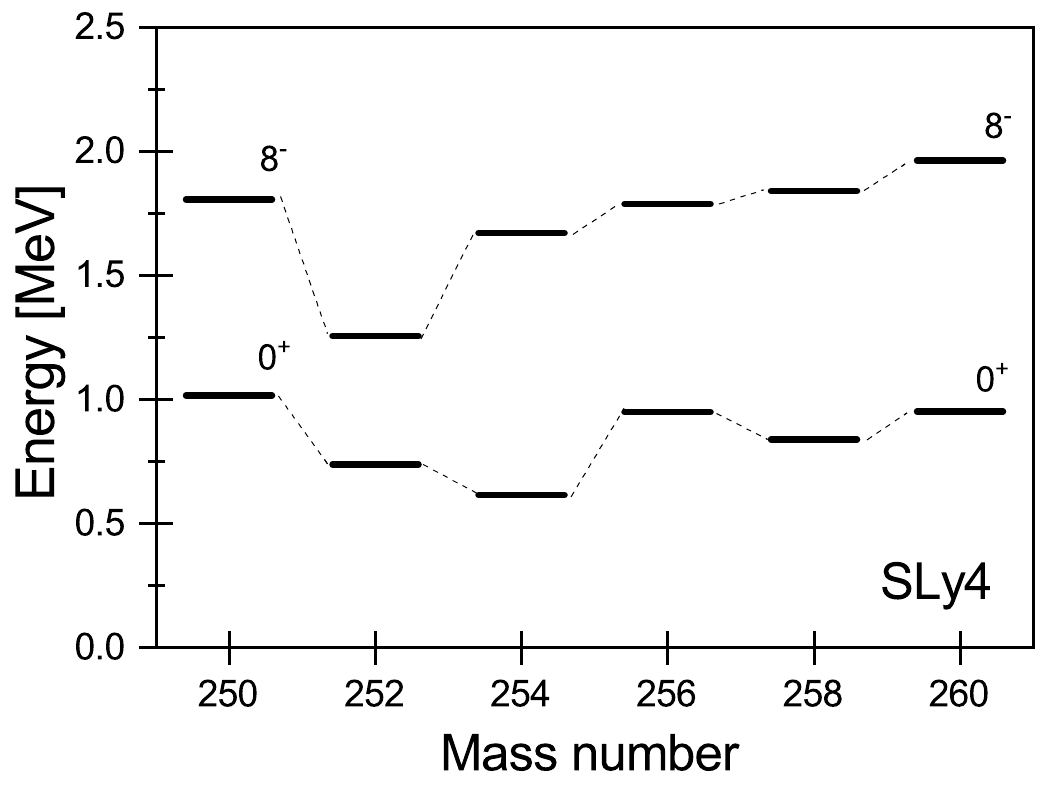}
\caption{The lowest QRPA 0$^+$ and 8$^-$ states in $^{250-260}$No.}
\label{0+8-}
\end{figure}

As seen from Table 2,
the experimental data demonstrate rather low-energies
for $8^-$ isomers in $^{252,254}$No. In accordance with this result, our calculations
also predict for these isotopes a  minimum
in the excitation energy of the isomers. This minimum can be explained by a drop of the neutron
pairing in $^{252,254}$No, mentioned above. The sequence of the calculated energies of the lowest
QRPA $8^-$  states is exhibited in Fig.\ref{0+8-}.

\begin{table*}[h!]
\begin{center}
\caption{Characteristics of the lowest QRPA $8^-$ states: QRPA energy $E$, reduced
probability $B(E98)$ for the transition from the ground state to the $8^-$ state,
main 2qp components $qq'$ of the QRPA state as well as their 2qp energies
$\epsilon_{qq'}$, single-particle energy differences $e_{qq'}$ and contributions to the state
norm $N_{qq'}$. The E$_{x}$ mark experiment excitation energies \cite{nndc}.}
\begin{tabular}{|c|c|c|c|c|c|c|c|}
\hline
      &  $E$  &  $B(E98)$ & $qq'$ & $\epsilon_{qq'}$  &  $e_{qq'}$ & $N_{qq'}$ & F-order \\
     &   (MeV) & (W.u.) &  & (MeV) &(MeV) & &  \\
\hline
$^{250}$No & 1.807 & 0.92  & $nn[624\downarrow, 734\uparrow]$  & 1.77  &0.91& 1.00 &  F+1,F+2\\
\hline
$^{252}$No & 1.257 & 1.07  & $nn[624\downarrow, 734\uparrow]$  & 1.21  &0.95& 1.00 & F, F+1 \\
& (E$_{x}$ = 1.254)&   &  &   &&  &  \\
\hline
$^{254}$No & 1.673 & 0.009 &  $nn[734\uparrow, 613\uparrow]$ & 1.70 &1.56& 1.00  &F,F+3 \\
& (E$_{x}$ = 1.295)&   &  &   &&  &  \\
\hline
$^{256}$No & 1.789 &  0.017  & $nn[734\uparrow, 613\uparrow]$  & 1.81  &1.57& 0.99 &  F-1,F+2\\
\hline
$^{258}$No & 1.842 & 1.80 &$ nn[734\uparrow, 613\uparrow]$ & 1.86 &1.52& 0.97 &F-2,F+1\\
\hline
$^{260}$No & 1.965 & 0.76 &$ pp[514\downarrow, 624\uparrow]$ & 1.90 &1.16& 1.00 &F+1,F+2\\
\hline
\end{tabular}
\end{center}
\label{tab:8-}
\end{table*}

\begin{table*}[h]
\begin{center}
\caption{The same as Table 2
but for the lowest excited $0^+$ states with reduced transition
probabilities $B(E20)$ and $\rho^2$(E0).}
\begin{tabular}{|c|c|c|c|c|c|c|c|}
\hline
      &  $E$  & $B(E20)$ &  $\rho^2(E0)$  & $qq'$ & $\epsilon_{qq'}$  & $N_{qq'}$ & F order \\
            &   (MeV) & (W.u.)&   (10$^{-3}$) &  & (MeV) & &  \\
\hline
$^{250}$No & 1.017 & 1.51 & 5.07  & $pp[514\downarrow,514\downarrow]$  & 1.12  & 0.47 &  F+1,F+1\\
                    &&           &                          &$pp[521\downarrow,521\downarrow]$ & 1.17 & 0.25 & F,F \\
\hline
$^{252}$No & 0.740 &0.18& 0.26  & $nn[734\uparrow,734\uparrow]$ & 1.07  & 0.53 &  F+1,F+1\\
                    &          & &                          &$nn[624\downarrow,624\downarrow]$ & 1.25 & 0.30 & F,F \\
\hline
$^{254}$No & 0.616 &0.06& 0.26 &  $nn[734\uparrow,734\uparrow]$ & 1.02 & 0.51  & F,F\\
                    &     (E$_{x}$=0.88)     & &                          &$nn[620\uparrow,620\uparrow]$ & 1.16 & 0.25 & F+1,F+1 \\
\hline
$^{256}$No & 0.950 & 1.03 &5.92  & $pp[514\downarrow,514\downarrow]$  & 1.10  & 0.33 & F+1,F+1 \\
                    &           &&                          &$nn[620\uparrow,620\uparrow]$ & 0.90 & 0.24 &  F,F\\
\hline
$^{258}$No & 0.840 &1.03 &19.54  &$nn[613\uparrow,613\uparrow]$ & 1.01  & 0.34 & F+1,F+1\\
                    &           & &                         &$nn[622\downarrow,622\downarrow]$ & 1.16 & 0.23 &  F,F\\
\hline
\hline
\end{tabular}
\end{center}
\label{tab:0+}
\end{table*}

\subsection{$0^+$ states}

In Table 3,
the features of the lowest excited QRPA  $0^+$ states in
$^{250-260}$No are presented. In particular, we show the reduced probability
$B(E20)$ for $E20$ transition $0^+0_{\rm gs} \to 2^+0_{\nu}$ (from the ground state to
the excited state $I^{\pi}K=2^+0_{\nu}$) and the normalized E0 transition probability
$\rho^2(E0)$ for $E0$ decay  $0^+0_{\nu} \to 0^+0_{\rm gs} $. The first ($\nu$ =1)
excited  QRPA monopole states  are considered.
The values $\rho^2(E0)$ are determined by monopole pairing vibrations
while $B(E20)$ values depend on the interplay of pairing and $\beta$-vibrations.
Note that low-energy pairing vibrations are well known in even-even deformed nuclei
in rare-earth and actinide regions~\cite{Sol76,SSS,Lo1,Lo2}. The mixed pairing
and $\beta$-vibrations provide the main and most natural mechanism of production of
low-energy $0^+$ excitations in  even-even deformed nuclei.

According to the Table 3,
the values $\rho^2(E0)$  and $B(E20)$ are decreased in
$^{252,254}$No. This is obviously caused by the drop of the neutron pairing in these two isotopes.
The drop also results in a similar behavior of the calculated
energies of $8^-$ and $0^+$ states,  shown in Fig.~\ref{0+8-}. The correlation
of excitation energies of $8^-$ and $0^+$ states is clearly seen: both them have
a pronounced minimum at A=252, 254. Fig.~\ref{0+8-} also shows that the
energies of $0^+$ states in $^{250-260}$No are systematically lower than the
energies of $8^-$ states. Besides, the $0^+$ states lie lower than other experimentally observed
non-rotational states in $^{252}$No ($E(2^-)$=0.929 MeV) and $^{254}$No ($E(3^+)$=0.987 MeV,
$E(4^+)$=1.203 MeV). Altogether, it seems that $0^+$ states are the lowest non-rotational
excitations in nobelium isotopes. Due to an admixture of the pairing vibrations, these states
can serve as a sensitive indicator of  the pairing.

Quite recently, a first non-rotational $0^+$ excitation in nobelium isotopes was observed
in $^{254}$No at 0.888 MeV~\cite{Forge,Forge2}. The state was treated as a result of coexistence
of  "normal" deformation $\beta \sim$0.3 and superdeformation $\beta \sim$1. The  former dominates
in the ground state and the latter  in the excited $0^+$ state. The treatment is done within a
simple model of two coupled states~\cite{Dela06}. Our SLy4 calculations also predict for $^{254}$No
the second deep minima in potential energy surface  at the superdeformation $\beta \sim$ 1.
However, as seen from Table 3,
the low-lying $0^+$ state can be obtained solely
at the "normal" deformation, i.e. without any admixture of the superdeformation.

Recent shell-model calculations~\cite{Dao} for $^{254}$No well reproduce
the measured energies of $K^\pi=8^-$ and $0^+$ states but the authors do not
discuss the origin of $0^+$ state. Altogether, our explanation of the 0.888–MeV state as a mixture of
pairing  and $\beta$-vibrations  seems to be the most plausible.

\section{Conclusions}

The analysis of the ground-state properties and features of excited $0^+$ and $8^-$ states
in even-even isotopes $^{250-260}$No was performed within fully self-consistent QRPA
method~\cite{Rep1,Rep2,Rep3} with SLy4 Skyrme force \cite{SLy4}. In particular, axial
quadrupole $\beta_2$ and hexadecapole $\beta_4$ deformations are determined. It is shown
that a deformation shell gap in the neutron single-particle spectra of $^{252,254}$No leads to
a significant decrease of the neutron pairing in these isotopes.

The calculation show that isomeric  $8^-$ states are represented by
neutron two-quasiparticle  configurations $nn[624\downarrow, 734\uparrow]$ in $^{252}$No
and  $nn[613\uparrow,734\uparrow]$ in $^{254-258}$No. We well reproduce
the energy of $8^-$ isomer in $^{252}$No but somewhat overestimate the energy of the isomer  in $^{254}$No.
In $^{252-258}$No, the low-energy $8^-$ isomers are produced by particle-hole transitions. Instead,
in $^{250,260}$No the $8^-$ transitions are of particle-particle character and so can hardly be observed.

The lowest excited $0^+$ states are interpreted as a mixture of pairing and $\beta$-vibrations.
These states can serve
as sensitive indicators of the pairing. Following our calculations, the first excited  $0^+$ states
should be the lowest non-rotational excitations in  $^{250-260}$No. The recently observed excited $0^+$
state in $^{254}$No
is treated as a superposition of the pairing and $\beta$-vibrations rather than the result of the
mixture of the  "normally" deformed and superdeformed shapes.

The energies of the lowest $0^+$ and $8^-$ excitations exhibit a minimum in  $^{252,254}$No, which can be
explained  by the drop of the neutron pairing in these isotopes.

\section{Acknowledgments}

We thank Prof. P.-G. Reinhard and Dr. A.Repko for providing codes SKYAX
and SKYAX-QRPA.


\end{document}